\documentclass[submission, Phys]{SciPost}

\usepackage{amsmath}
\usepackage{amssymb}
\usepackage{esint}
\usepackage{graphicx}
\usepackage{bbold}
\usepackage[normalem]{ulem}

\usepackage{setspace}

\usepackage{color}

\hypersetup{
    colorlinks,
    linkcolor={red!50!black},
    citecolor={blue!50!black},
    urlcolor={blue!80!black}
}

\usepackage{subfig}
\usepackage{soul}
\usepackage[normalem]{ulem}

\usepackage[bitstream-charter]{mathdesign}
\DeclareSymbolFont{usualmathcal}{OMS}{cmsy}{m}{n}
\DeclareSymbolFontAlphabet{\mathcal}{usualmathcal}

\definecolor{violet(colorwheel)}{rgb}{0.0, 0.7, 1}

\usepackage{breakurl}
\usepackage{seqsplit}

\makeatletter

\begin{document}


\begin{center}{\Large \textbf{
Quantum many-body thermal machines enabled by atom-atom correlations
}}\end{center}

\begin{center}
R. S. Watson\textsuperscript{1  $\dagger$} and
K. V. Kheruntsyan\textsuperscript{1$\star$}
\end{center}

\begin{center}
{\bf 1}  School of Mathematics and Physics, University of Queensland, Brisbane, Queensland 4072, Australia
\\

${}^\dagger${\small \sf raymon.watson@uq.edu.au} \\
${}^\star$ {\small \sf karen.kheruntsyan@uq.edu.au}
\end{center}

\date{\today}

\section*{Abstract}
{\bf
Particle-particle correlations, characterized by Glauber's second-order correlation function,
play an important role in the understanding of various phenomena in radio and optical astronomy, quantum and atom optics, particle physics, condensed matter physics, and quantum many-body theory. However, the relevance of such correlations to quantum thermodynamics has so far remained illusive. Here, we propose and investigate a class of quantum many-body thermal machines whose operation is directly enabled by second-order atom-atom correlations in an ultracold atomic gas. More specifically, we study quantum thermal machines that operate in a sudden interaction-quench Otto cycle and utilize a one-dimensional Lieb-Liniger gas of repulsively interacting bosons as the working fluid. 
The atom-atom correlations in such a gas are different to those of a classical ideal gas, and are a result of the interplay between interparticle interactions, quantum statistics, and thermal fluctuations. 
We show that operating these thermal machines in the intended regimes, such as a heat engine, refrigerator, thermal accelerator, or  heater, would be impossible without such atom-atom correlations. Our results constitute a step forward in the design of conceptually new quantum thermodynamic devices which take advantage of uniquely quantum resources such as quantum coherence, correlations, and entanglement. 
}


\vspace{10pt}
\noindent\rule{\textwidth}{1pt}
\tableofcontents\thispagestyle{fancy}
\noindent\rule{\textwidth}{1pt}
\vspace{10pt}

\section{Introduction}
\label{sec:into}

Quantum thermal machines (QTM), such as quantum heat engines (QHE), refrigerators, and quantum batteries, are central in the theoretical and experimental development of the emerging field of quantum thermodynamics \cite{vinjanampathy2016quantum,kosfloff2014quantum,potts2024quantum,Bhattacharjee2021,Gluza2021}. Their primary utility is to explore the fundamental laws of thermodynamics in the quantum realm and to  demonstrate possible advantages gained by utilising quantum resources. Accordingly, QTM's are expected to play a similar role in the development of quantum technologies as classical heat engines played in fostering scientific advances during
the Industrial Revolution. In the past decade, progress in the control over quantum platforms, such as single ions \cite{Rossnagel2016single,Ca-ion-spin-engine,Maslennikov2019,VanHorne2020}, nitrogen vacancy centers \cite{Nitrogen-vacancy-heat-engine}, and single-atom impurities immersed in an ultra-cold atomic bath \cite{bouton2021quantum}, have led to the realization of single-particle QHE's and quantum refrigerators. Such single-particle QTM's represent the ultimate limit in the realization of an `infinitesimal machine' \cite{feynman2018there}.

However, to utilize the breadth of quantum resources available, one must move beyond single-particle systems---to engines that utilize interacting many-particle systems. Such QHE's are uniquely positioned to take advantage of quantum resources, such as entanglement \cite{brandao2008entanglement,funo2013thermo,alicki2013entanglement,hilt2009system}, correlations \cite{oppenheim2002thermo,llobet2015extractable,Huber_2015}, or quantum coherence \cite{narasimhachar2015low,Korzekwa2016extraction,
lostaglio2015description,bernardo2020unraveling,kammerlander2016coherence,Cwiklinski2015limitations}, to enhance the performance of classical heat engines \cite{jaramillo2016quantum} or perform entirely new tasks that would be impossible classically \cite{halpern2019quantum}.
In particular, control over inter-particle interactions allows for the creation of  strictly many-body QHE's \cite{jaramillo2016quantum,mathieu2016scaling,li2018efficient,chen2019interaction,keller2020feshbach,Boubakour_2023,Nautiyal_2024,Williamson_2024,Williamson_2025},
which have recently been realized in the laboratory \cite{koch2022making,simmons2023thermodynamic}. 
These very recent experimental developments underscore the need for further studies of thermodynamic processes in the context of interacting quantum many-body thermal machines.

Here, we propose a quantum many-body Otto heat engine---as well as related thermal machines including the Otto refrigerator, thermal accelerator, and heater---using a uniform one-dimensional (1D) Bose gas with repulsive contact interactions as the working fluid. In the proposed Otto cycles, the unitary work strokes are driven by a sudden quench of the interaction strength. We demonstrate how the thermodynamic performance, in particular net work and efficiency, of these many-body QTM's can be calculated through the experimentally measurable atom-atom local pair correlation \cite{kheruntsyan2003pair,kheruntsyan2005finite,kinoshita2005local}. 
The atom-atom local correlation, $g^{(2)}(0)$, is described by the second-order Glauber correlation function $g^{(2)}(r)$ at zero interparticle separation (i.e., when $r=0$, where $r=|x-x'|$ is the distance between the two particles with positions $x$ and $x'$) and characterizes the probability of pairs of atoms to be found at the same point, relative to uncorrelated atoms.

The benefits of using the 1D Bose gas as the working fluid in the proposed Otto cycles is that the underlying theoretical model---the Lieb-Liniger model---is exactly solvable in the uniform limit via the Yang-Yang thermodynamic Bethe ansatz (TBA) \cite{liebliniger,yang1969thermodynamics,korepin1993}, in addition to being experimentally realizable using ultracold atomic gases confined to highly anisotropic traps 
\cite{gorlitz2001realization,greiner2001phase,schreck2001quasipure}.
This offers unique opportunities for gaining physical insights into the performance of such Otto QTM's as a tractable and testable quantum many-body problem.
Additionally, the Lieb-Liniger gas has a rich phase diagram spanning several nontrivial regimes \cite{kheruntsyan2003pair,kheruntsyan2005finite}, from the weakly interacting quasicondensate \cite{Petrov_2000_Regimes,Mora-Castin-2003} through to the strongly interacting Tonks-Girardeau regime of fermionization \cite{Girardeau_1960,Kinoshita1125}.
The atom-atom correlation within these regimes takes on a wide range of values between $0<g^{(2)}(0)<2$ \cite{kheruntsyan2003pair,kheruntsyan2005finite,Sykes2008}, depending on the temperature and interaction strength, which aids the operation of the proposed Otto cycles under a variety of conditions.
We evaluate the performance of the 1D Bose gas Otto QTM's, but we emphasise that the broad conclusions arrived at here are not limited to the Lieb-Liniger model.

A related interaction-driven Otto engine cycle with a uniform 1D Bose gas as the working fluid was previously studied by Chen \textit{et al.} in Ref.~\cite{chen2019interaction} in the opposite limit of a quasi-static, or isentropic (rather than sudden quench), work strokes. In this case, the performance of the engine cannot be expressed merely in terms of Glauber's second-order correlation function, as both the kinetic and interaction energies evolve and change during the interaction quench. Nevertheless, the performance of such an isentropic Otto engine could still be evaluated analytically in the low-temperature regime, owing to the known thermodynamic properties of the system using the Tomonaga-Luttinger liquid approach.

\section{Interaction-driven Otto heat engine.}
\label{sec:Otto}

In a uniform 1D Bose gas, described by the integrable Lieb-Liniger model \cite{liebliniger} (see Appendix \ref{appendix:LL}), the interatomic interaction strength $\chi$ can be expressed in terms of the harmonic trap frequency $\omega_{\perp}$ in the tightly confined (transverse) direction and the 3D $s$-wave scattering length $a_s$ as $\chi \simeq 2 \hbar \omega_\perp a_s$ \cite{Olshanii1:998}. Accordingly, changing the interaction strength $\chi$ may be achieved by either tuning the external trapping potential that controls the transverse confinement $\omega_{\perp}$ or by changing the scattering length $a_s$ by means of a magnetic Feshbach resonance \cite{chin2010feshbach}. The former option leads to a volume change of the gas (i.e. transverse expansion or compression), and hence can be thought of as analogous to mechanical work in the conventional Otto cycle. However, changing $\chi$ via a change of the scattering length leads to identical results, which then justifies our referral to the engine cycle as the Otto cycle \cite{Kosloff1984quantum,Kosloff2017quantum,li2018efficient,chen2019interaction,koch2022making,bouton2021quantum,zheng2014power,watanabe2017cycles,Abah_2017,Myers2022boosting} regardless of the means of tuning the interaction strength (for further discussion, see (see Appendix \ref{appendix:Otto})).
We emphasize, however, that the dynamics of the quantum Otto cycle explored here are strictly longitudinal, with the gas always remaining in its transverse ground state.

\begin{figure}[t!]
    \centering
 \includegraphics[width=12cm]{ 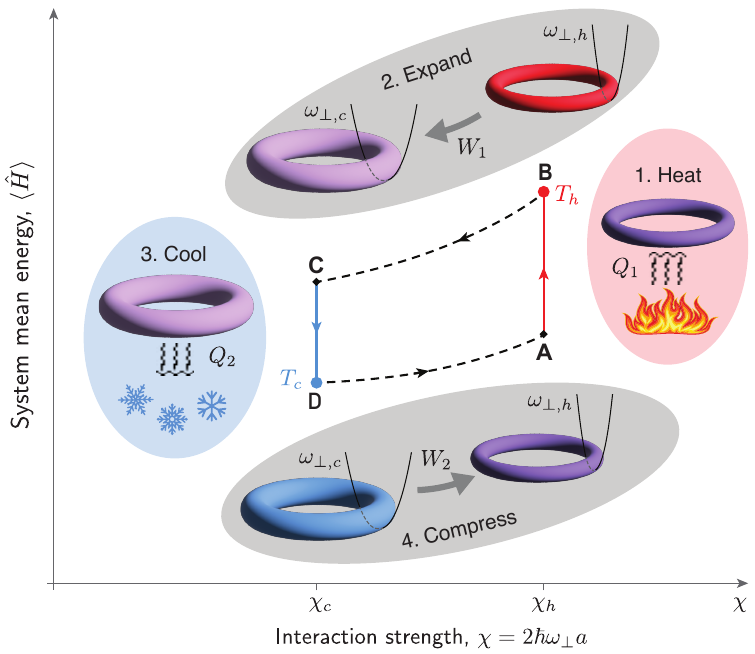}   
   \caption{An interaction-driven quantum many-body Otto engine cycle, operating between two interaction strengths, $\chi_{c}$ and $\chi_{h}$. Unitary work strokes (\textbf{B}-\textbf{C} and \textbf{D}-\textbf{A}) are shown in dashed black, while non-unitary thermalization strokes (\textbf{A}-\textbf{B} and \textbf{C}-\textbf{D}) are color-coded to the cold (blue) and hot (red) reservoirs at temperatures $T_{c}$ and $T_{h}$, respectively. We further note that the work strokes \textbf{B}-\textbf{C} and \textbf{D}-\textbf{A} are shown as dashed lines to reflect the fact that the system does not continuously pass through these intermediate points as in a quasi-static process of a typical engine; in fact, there is no way to represent the passage from \textbf{B} to \textbf{C} and from \textbf{D} to \textbf{A} by some continuous lines in this sudden (instantaneous) interaction quench engine, where only the differences in the energies of the system at the end points of work strokes can be defined and calculated. Nevertheless we use this standard Otto engine diagram for illustrative purposes only -- to explain the four strokes of our engine as if it was operating quasi-statically.}
  \label{fig:Engine_Diagram}
\end{figure}

The interaction-driven Otto engine cycle, which we thus consider, consists of four strokes (see Fig.~\ref{fig:Engine_Diagram}): 
\begin{itemize}
\vspace{-0.2cm}
\item[(1)] \textit{Thermalization with hot reservoir}, \textbf{A}$\to$\textbf B: the working fluid consisting of $N$ total atoms at interaction strength $\chi_{h}$, is connected to a hot ($h$) reservoir at temperature $T_{h}$. The working fluid begins in some non-equilibrium state $\textbf{A}$ (corresponding to the final state of the previous stroke) and is left to equilibrate with the reservoir; it takes in heat $Q_1\!=\!\langle \hat{H}\rangle_{\textbf{B}} \!-\!\langle \hat{H}\rangle_{\textbf{A}}\!>\!0$, which is to be partially converted into beneficial work in the subsequent stroke. Here $\hat{H}$ is the system Hamiltonian (see Appendix \ref{appendix:LL}), and $\langle \hat{H}\rangle_{\textbf{j}}$ is its expectation value, i.e., the total energy of the system, in state $\textbf{j}=\{\textbf{A,B,C,D}\}$.
\vspace{-0.3cm}
\item[(2)] \textit{Unitary expansion, \bf{B}$\to$\bf{C}}: the working fluid, now in a thermal equilibrium state at temperature $T_{h}$, is decoupled from the hot reservoir and has its interaction strength quenched from $\chi_{h}$ to $\chi_{c}\!<\!\chi_{h}$, generating beneficial 
work $W_1\!=\!\langle \hat{H}\rangle_{\textbf{C}} \!-\!\langle \hat{H}\rangle_{\textbf{B}}\!<\!0$ done by the fluid. 
\vspace{-0.3cm}
\item[(3)] \textit{Thermalization with cold reservoir}, \textbf{C}$\to$\textbf{D}: the working fluid is connected to a cold ($c$) reservoir at temperature $T_{c}<T_{h}$ and allowed to equilibrate at constant interaction strength $\chi_{c}$ while dumping energy in the form of heat $Q_2\!=\!\langle \hat{H}\rangle_{\textbf{D}} \!-\!\langle \hat{H}\rangle_{\textbf{C}}\!<\!0$ into the cold reservoir. 
\vspace{-0.3cm}
\item[(4)] \textit{Unitary compression}, \textbf{D}$\to$\textbf{A}: disconnected from the cold reservoir, the working fluid has its interaction strength quenched from $\chi_{c}\!\to\!\chi_{h}$, with work $W_2\!=\!\langle \hat{H}\rangle_{\textbf{A}} \!-\!\langle \hat{H}\rangle_{\textbf{D}}\!>\!0$ done on the fluid, and the system returning to the initial state of the overall cycle.
\end{itemize}
\vspace{-0.3cm}
Such an engine cycle generates net beneficial work (done by the fluid) if the total work $W \!=\! W_1 \!+\! W_2\!<\!0$, i.e., if $|W_1|>W_2$ (or $Q_1>|Q_2|$), with efficiency 
\begin{equation}
\eta = -\frac{W}{Q_1} = 1 - \frac{|Q_2|}{Q_1}, 
\label{eq:efficiency_def}
\end{equation}
where we used the conservation of energy $W+Q=0$, with $Q=Q_1 + Q_2$ being the total heat \cite{CallenHerbertB1985Taai}.

\section{Work from second-order Glauber correlations.}
\label{sec:g2}

In this work, we specifically consider a sudden or instantaneous quench of the interaction strength $\chi$ in the unitary strokes (2) and (4). (For a discussion of ``instantaneity'' of the sudden quench, see Appendix \ref{appendix:instantaneity}). 
Under a sudden interaction quench, the initial ($i$) and final ($f$) expectation values over field operators in the system Hamiltonian, i.e., the expectation values before and after the quench,
remain unchanged as the system did not have sufficient time to evolve into a new state. Hence, the only contribution to the difference in total energy before and after the quench, $ \langle\hat{H} \rangle_f - \langle \hat{H} \rangle_i$, comes from the difference between the interaction terms, $\frac{1}{2}\chi_f\int dz \langle \hat{\Psi}^{\dagger}\hat{\Psi}^{\dagger} \hat{\Psi}\hat{\Psi} \rangle_f-\frac{1}{2}\chi_i\int dz \langle \hat{\Psi}^{\dagger}\hat{\Psi}^{\dagger} \hat{\Psi}\hat{\Psi} \rangle_i$, where $\langle \hat{\Psi}^{\dagger}\hat{\Psi}^{\dagger} \hat{\Psi}\hat{\Psi} \rangle_f=\langle \hat{\Psi}^{\dagger}\hat{\Psi}^{\dagger} \hat{\Psi}\hat{\Psi} \rangle_i$ in a sudden quench, and $\hat{\Psi}^{\dagger}(z)$ and $\hat{\Psi}(z)$ represent the field creation and annihilation operators. Accordingly, the energy difference can be expressed as $ \langle\hat{H} \rangle_f - \langle \hat{H} \rangle_i\!=\!  \frac{1}{2}(\chi_f \!-\! \chi_i)\overline{G^{(2)}_i}$,
where we have defined the total (integrated) second-order correlation of the thermal equilibrium state  \cite{kheruntsyan2005finite}: 
 \begin{equation}
 \overline{G^{(2)}_i} \!=\! \int dz \langle \hat{\Psi}^\dagger \hat{\Psi}^\dagger \hat{\Psi} \hat{\Psi} \rangle_i.
 \end{equation}

Identifying the $i$ and $f$ states as points \textbf{B} (hot, $h$) and \textbf{C}, or as \textbf{D} (cold, $c$) and \textbf{A} in the digram of Fig. \ref{fig:Engine_Diagram}, the net work of the Otto heat engine  can be expressed as 
\begin{equation}\label{eq:Work}
    W = -\frac{1}{2}(\chi_{h} - \chi_{c})\left( \overline{G^{(2)}_{h}} - \overline{G^{(2)}_{c}}\right),
\end{equation}
whereas the efficiency, Eq.~\eqref{eq:efficiency_def}, of the engine as
\begin{equation}\label{eq:Efficiency}
    \eta = 1 - \frac{ \langle\hat{H} \rangle_{h} - \langle \hat{H} \rangle_{c} - \frac{1}{2}\left(\chi_{h} - \chi_{c} \right) \overline{G^{(2)}_{h}}}{ \langle\hat{H} \rangle_{h} - \langle \hat{H} \rangle_{c}   - \frac{1}{2}\left(\chi_{h} - \chi_{c} \right)  \overline{G^{(2)}_{c}}}.
\end{equation}
These equations allow for evaluation of the performance of the interaction-driven Otto engine under a sudden quench through solely the \emph{equilibrium} properties of the gas as all relevant expectation values here are with respect to $h$ (\textbf{B}) or $c$ (\textbf{D}) states.

For a uniform 1D Bose gas, the total correlation in hot or cold thermal equilibrium state may be expressed as $\overline{G^{(2)}_{h(c)}} \!=\!N \rho g^{(2)}_{h(c)}(0)$, where $g^{(2)}_{h(c)}(0)$ is the normalized same-point pair correlation function also referred to as Glauber second-order correlation
(see Appendix \ref{appendix:Glauber}). Combining this with Eq.~\eqref{eq:Work}, the net work per particle can be expressed as
\begin{equation}\label{eq:Work_Uniform}
    \frac{W}{N} = -\frac{\hbar^2 \rho^2}{ 2 m}(\gamma_{h} - \gamma_{c})\left( g^{(2)}_{h}(0) - g^{(2)}_{c}(0) \right),
\end{equation} 
meaning the net work is directly proportional to the difference between atom-atom correlations of the 1D Bose gas in the hot and cold thermal equilibrium states. This simple relationship between thermodynamic work and Glauber second-order correlation function, $g^{(2)}_{h(c)}(0)$, at thermal equilibrium represents the key result of this work.

\begin{figure}[t!]
    \centering
 \includegraphics[width=10cm]{ 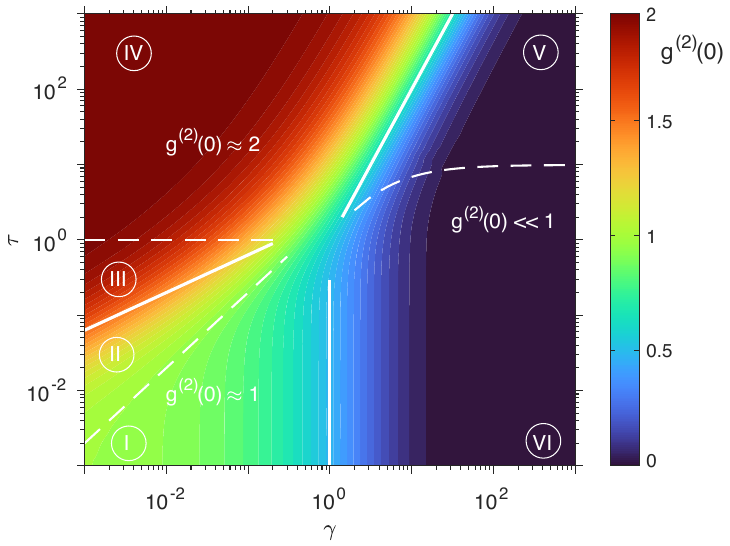}   
   \caption{Atom-atom pair correlations, described by Glauber's $g^{(2)}(0)$ correlation function, for the uniform 1D Bose gas parametrized via the dimensionless interaction strength $\gamma$ and dimensionless temperature $\tau$, and evaluated using the exact Yang-Yang TBA \cite{kheruntsyan2003pair,yang1969thermodynamics}. The pair correlation can be also calculated analytically in six asymptotic regimes labeled via I-VI; the solid and dashed lines show the crossover boundaries between different asymptotic regimes and sub-regimes (see text and Appendix  \ref{appendix:Glauber}  for details).}
\label{fig:regimes}
\end{figure}

Before continuing the discussion, we momentarily pause here to comment on the following aspect of our work strokes. As we have seen above, the work in the unitary strokes \textbf{B}-\textbf{C} and \textbf{D}-\textbf{A} can be evaluated simply from energy differences between the end points. While this is consistent with the notion that work is not an observable \cite{Talkner2007} as it requires the use of two-time measurement approach (before and after the quench), the fact that it depends only on energy differences is not typical in other engines. More generally, the amount of work done by the working fluid depends on the specific path taken, not just the initial and final states. Here, however, the ``specific'' path taken is defined by the sudden interaction quench itself, and therefore no contradiction arises.

In order to evaluate the net work using Eq.~\eqref{eq:Work_Uniform}, we note that the pair correlation function $g^{(2)}(0)$ is a thermodynamic quantity that can be calculated numerically using exact TBA, as well as analytically via approximate methods in six asymptotic regimes \cite{kheruntsyan2003pair,kerr2023analytic} parametrized in terms of the dimensionless interaction strength, $\gamma \!=\! m \chi/\hbar^2 \rho$, and dimensionless temperature, $\tau \!=\! 2 m k_B T/\hbar^2 \rho^2$, where $m$ is the boson mass and $\rho\!=\!N/L$ is the 1D density for $N$ atoms in a system of length $L$ . These asymptotic regimes (see Appendix \ref{appendix:Glauber} for details) can be used to characterize the equilibrium phase diagram of the uniform 1D Bose gas, all connected by smooth crossovers and shown in  Fig.~\ref{fig:regimes}, where we denote them via I-VI. Physically these different regimes correspond to \cite{kheruntsyan2003pair,kerr2023analytic} :
\begin{itemize}
\vspace{-0.2cm}
\item[(I):] Weakly interacting ($\gamma\ll1$) quasi-condensate described by the Bogoliubov theory and dominated by quantum fluctuations ($\tau\ll2\gamma$);
\vspace{-0.3cm}
\item[(II):] Weakly interacting ($\gamma\ll1$) quasi-condensate described by the Bogoliubov theory and dominated by thermal fluctuations ($2\gamma\ll \tau \ll 2\sqrt{\gamma}$);
\vspace{-0.3cm}
\item[(III):] Highly-degenerate ($\tau\ll1$) nearly ideal Bose gas that can be described by perturbation theory in $\gamma$ ($2\sqrt{\gamma}\ll \tau $); 
\vspace{-0.3cm}
\item[(IV):] Non-degenerate ($\tau\gg 1$) nearly ideal Bose gas that can be described by perturbation theory in $\gamma$ ($\tau\gg\max\{1,\gamma^2\}$);
\vspace{-0.3cm}
\item[(V):] Strongly interacting ($\gamma\!\gg\!1$) or Tonks-Girardeau regime of high-temperature fermionization that can be described by perturbation theory in $1/\gamma$ ($\pi^2/(1+2/\gamma)^2\!\ll\!\tau\!\ll\! \gamma^2$);
\vspace{-0.3cm}
\item[(VI):] Strongly interacting ($\gamma\gg1$) or Tonks-Girardeau regime of low-temperature fermionization that can be described by perturbation theory in $1/\gamma$ ($\tau\ll \pi^2/(1+2/\gamma)^2$). 
\end{itemize}

From Eq.~\eqref{eq:Work_Uniform}, and given that $\gamma_{h}$ is always larger than $\gamma_{c}$ (as $\chi_h>\chi_c$; see Fig.~\ref{fig:Engine_Diagram}), we see that if the local pair correlations did not depend on the respective interaction strengths and temperatures, i.e. if they were the same, $ g^{(2)}_{h}(0)\! =\! g^{(2)}_{c}(0)$, then the net work per particle would vanish. 
We therefore conclude that extracting net work ($W\!<\!0$) from this Otto cycle, and hence operating it as a heat engine, can only be enabled by atom-atom correlations; more specifically, the only way to extract net work is to have $ g^{(2)}_{h}(0)\!>\!g^{(2)}_{c}(0)$ 
(see Fig. \ref{fig:g2_parameter_space} \,(c) in Appendix \ref{appendix:Glauber} for an illustration).

\begin{figure}[t!]
  \centering
   \includegraphics[width=12cm]{ 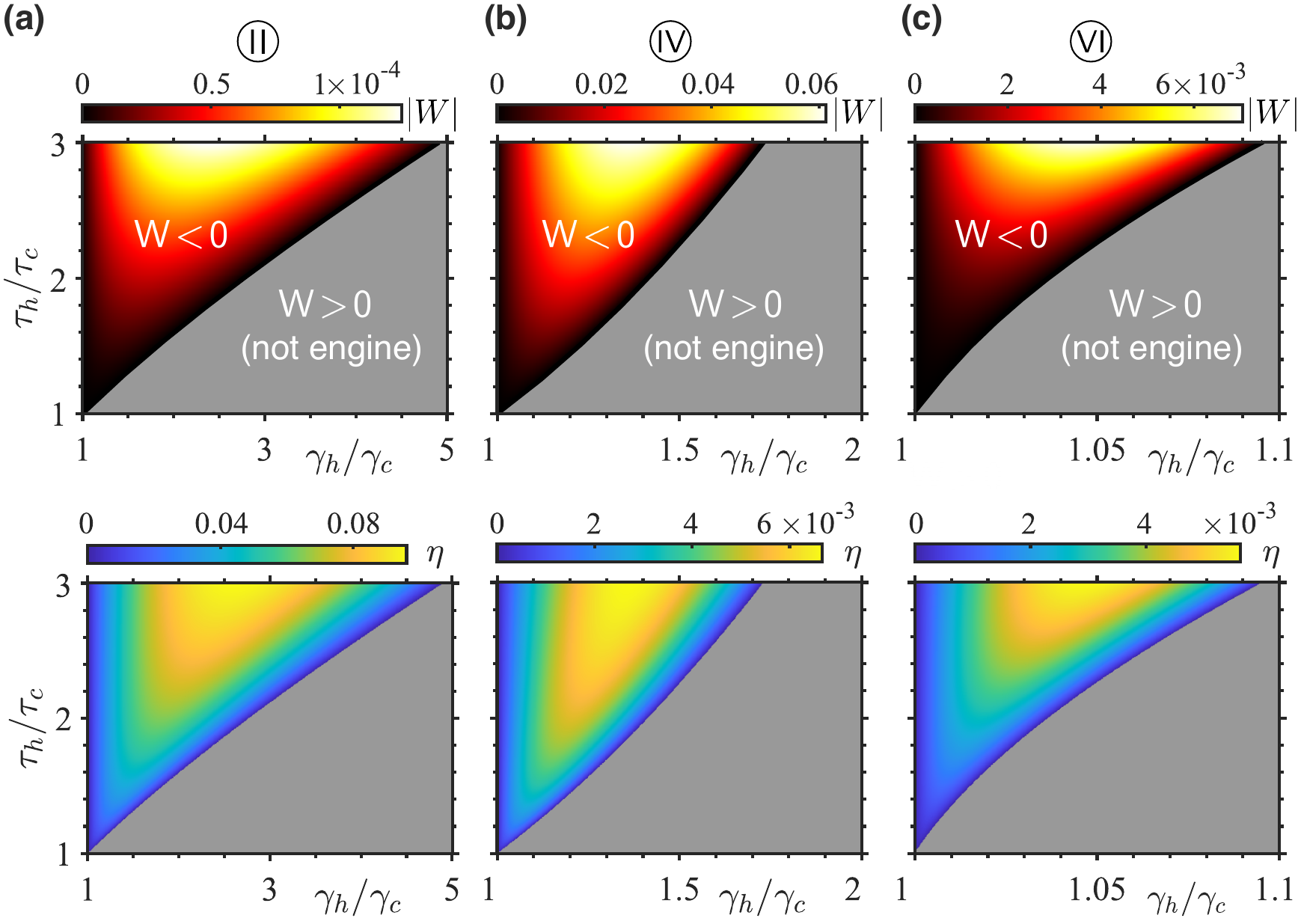}   
   \caption{Performance of the interaction driven quantum Otto heat engine. Columns (a)--(c)  demonstrate net work, $|W|$, and efficiency, $\eta$, as a function of the ratio of interaction strength, $\gamma_h/\gamma_c$, and temperature, $\tau_h/\tau_c$, between the hot ($h$) and cold ($c$) thermal equilibrium states. Regions corresponding to $W\!>\!0$, here colored grey, are outside the engine operation regime and are considered further in Figs.~\ref{fig:operational_regimes} and \ref{fig:CoP}  below. The example of panel  (a) is for $\gamma_c\!=\!10^{-3}$ and $\tau_c\!=\!10^{-2}$, where $\gamma_h/\gamma_c$ and $\tau_h/\tau_c$ explores the parameter range within the region II of the equilibrium phase diagram of Fig. \ref{fig:regimes}. 
   Similarly, panel (b) explores region IV, with $\gamma_c\!=\!1$ and $\tau\!=\!10$, whereas (c) explores region VI, with $\gamma_c\!=\!10$ and $\tau\!=\!1$.
   }
  \label{fig:engine_performance}
\end{figure}

Net work, Eq.~\eqref{eq:Work_Uniform}, and efficiency, Eq.~\eqref{eq:Efficiency}, of this quantum Otto engine, calculated for simplicity using analytic approximations to the atom-atom correlation function and total energy \cite{kheruntsyan2003pair,DeRosi2022anomaly}, are shown in Fig.~\ref{fig:engine_performance} as a function of the ratio of interaction strengths, $\gamma_{h}/\gamma_{c}$, and temperatures, $\tau_{h}/\tau_{c}$, for three of the six asymptotic regimes (for results outside of the regimes of analytic approximations, see Appendix \ref{appendix:TBA}).
Notably, in this Otto engine cycle, for any fixed value of the temperature ratio, the interaction strength quench corresponding to maximum net work is approximately the same as that providing maximum efficiency; this occurs as, to first order, the heat intake $Q_1$ varies slowly with $\gamma_h/\gamma_c$, meaning $\eta \!\propto\! W$ (see Appendix \ref{appendix:max-efficiency}).
The observed increase of net work and efficiency in all regimes under large temperature ratio may be attributed 
to the fact that the local correlation of the hot thermal state in Eq.~\eqref{eq:Work_Uniform} is always a monotonically increasing function of $\tau$.
However, as the correlation function is also monotonically decreasing under $\gamma_h$, this results in no extractable net work under sufficiently large interaction strength ratios for any given temperature ratio.

At a glance, one may conclude that the net work, which is enabled through the $g^{(2)}(0)$ correlation function, is maximized under the largest possible difference in correlation function, i.e. $g^{(2)}_h(0)\!-\!g^{(2)}_c(0)\!\simeq\! 2$.
However, to achieve this, while also guaranteeing that $\gamma_h \!>\! \gamma_c$, would require an unrealistically high (from practical point of view) temperature ratio to operate between regimes VI and IV, shown in Fig. \ref{fig:regimes}.
Rather, we observe that, while the $g^{(2)}(0)$ correlation function is responsible for \emph{enabling} operation as a heat engine, the \emph{magnitude} of net work is governed more strongly by the difference in the interaction strengths, $\gamma_h-\gamma_c$, which is more susceptible to large variations in practice using, e.g., magnetic Feshbach tuning of the $s$-wave scattering length. 

Consequently, it is in the weakly interacting ($\gamma\ll1$) region II, shown in Fig.~\ref{fig:engine_performance}\,(a), where $\gamma_h-\gamma_c$ is small, that we observe the lowest magnitude of net work. In comparison, the magnitude of net work is largest in regime IV, shown in Fig.~\ref{fig:engine_performance}\,(b), where $\gamma\!\sim\!1$ and hence the difference $\gamma_h\!-\!\gamma_c$ can also be on the order of one. The same considerations apply to the strongly interacting ($\gamma\!\gg\!1$) regime VI, where one can operate under the largest magnitudes of interaction strengths; however, in this regime the net work is diminished due to the vanishing of correlation itself ($g^{(2)}\!\ll\!1$) arising from the effect of fermionization \cite{Girardeau_1960}. In contrast to these observations, the efficiency of the engine, Eq.~\eqref{eq:Efficiency}, is inversely dependent on the total energy of the thermal states, which is minimal in the weakly interacting low temperature regime II, which thus has the largest efficiency.
Comparison of this sudden quench Otto engine cycle against the maximum possible efficiency cycle, i.e. a quasi-static isentropic quench, is discussed in Appendix \ref{appendix:isentropic-quench}.

\section{Interaction-driven Otto accelerator, heater, and refrigerator.}
\label{sec:beyond}

Glauber's $g^{(2)}(0)$ correlation function is inherently dependent on the interaction strength and temperature \cite{kheruntsyan2003pair}. This implies that the condition for the Otto cycle to operate as a heat engine, $g^{(2)}_c(0)\!<\!g^{(2)}_h(0)$, where $\gamma_c\!<\!\gamma_h$, cannot hold under large quenches of interaction strength as the gas becomes increasingly fermionized (i.e., $g_h^{(2)}(0)\!\to\!0$) in the limit $\gamma \! \to \! \infty$ \cite{Girardeau_1960}.

However, beyond the heat engine [E] operation regime, a further three QTM's are thermodynamically allowed \cite{buffoni2019quantum}, namely, the accelerator [A], heater [H], and refrigerator [R]. For these QTM's, one may define a coefficient of performance (CoP) according to \cite{SchroederD_ThermalPhysics}:
\begin{equation}
    \mathrm{CoP[QTM]} = \frac{\mathrm{benefit\,of\,operation}}{\mathrm{cost\,of\,operation}}.
    \label{eq:CoP}
\end{equation}
We now discuss the operating conditions and the CoP's of these three additional QTM's in detail.

\subsection{Accelerator [A]}

\begin{figure}[t!]
      \centering
\includegraphics[width=12cm]{ 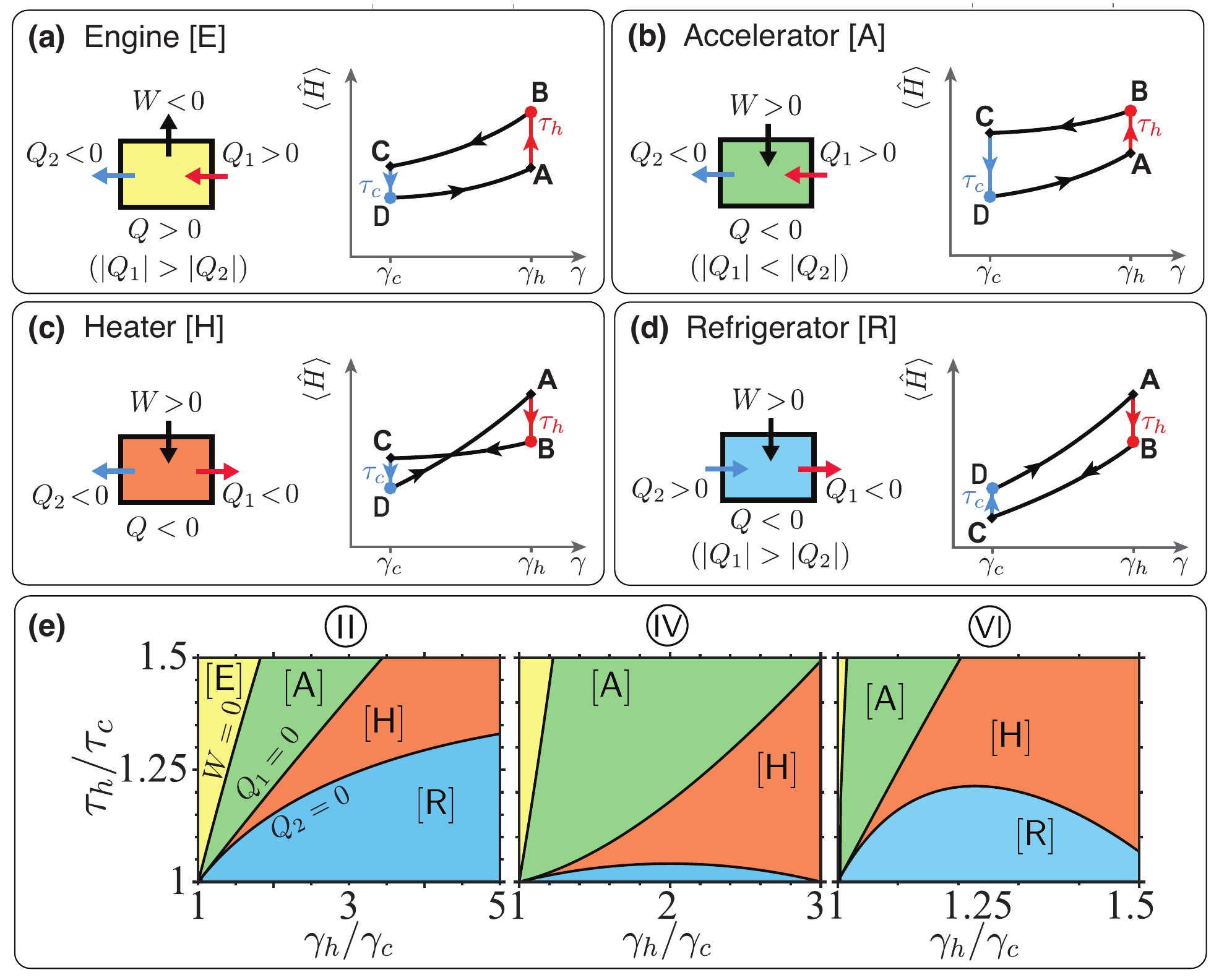}
    \caption{Energy flow diagrams and schematics of the interaction driven Otto cycles in different regimes: (a) heat engine [E], (b) accelerator [A], (c) heater [H], and (d) refrigerator [R]. In all  four regimes, the difference between the energies of the hot and cold thermal state (\textbf{B} and \textbf{D}) are the same, however the energies of the resulting state after the interaction driven work strokes can be different, leading to these four different physically valid outcomes. Panel (d) shows the layout of how the operating regimes of these different QTM's depend on the ratio of interaction strengths, $\gamma_h/\gamma_c$, and temperatures, $\tau_h/\tau_c$, of the hot and cold thermal states in same the three asymptotic regimes (II, IV, and VI) as in Fig.~\ref{fig:engine_performance}. The respective coefficients of performance of these QTM's within these operating boundaries are shown in Fig.~\ref{fig:CoP}.
    }
\label{fig:operational_regimes}
\end{figure}

\begin{figure}[tbp]
      \centering
\includegraphics[width=12cm]{ 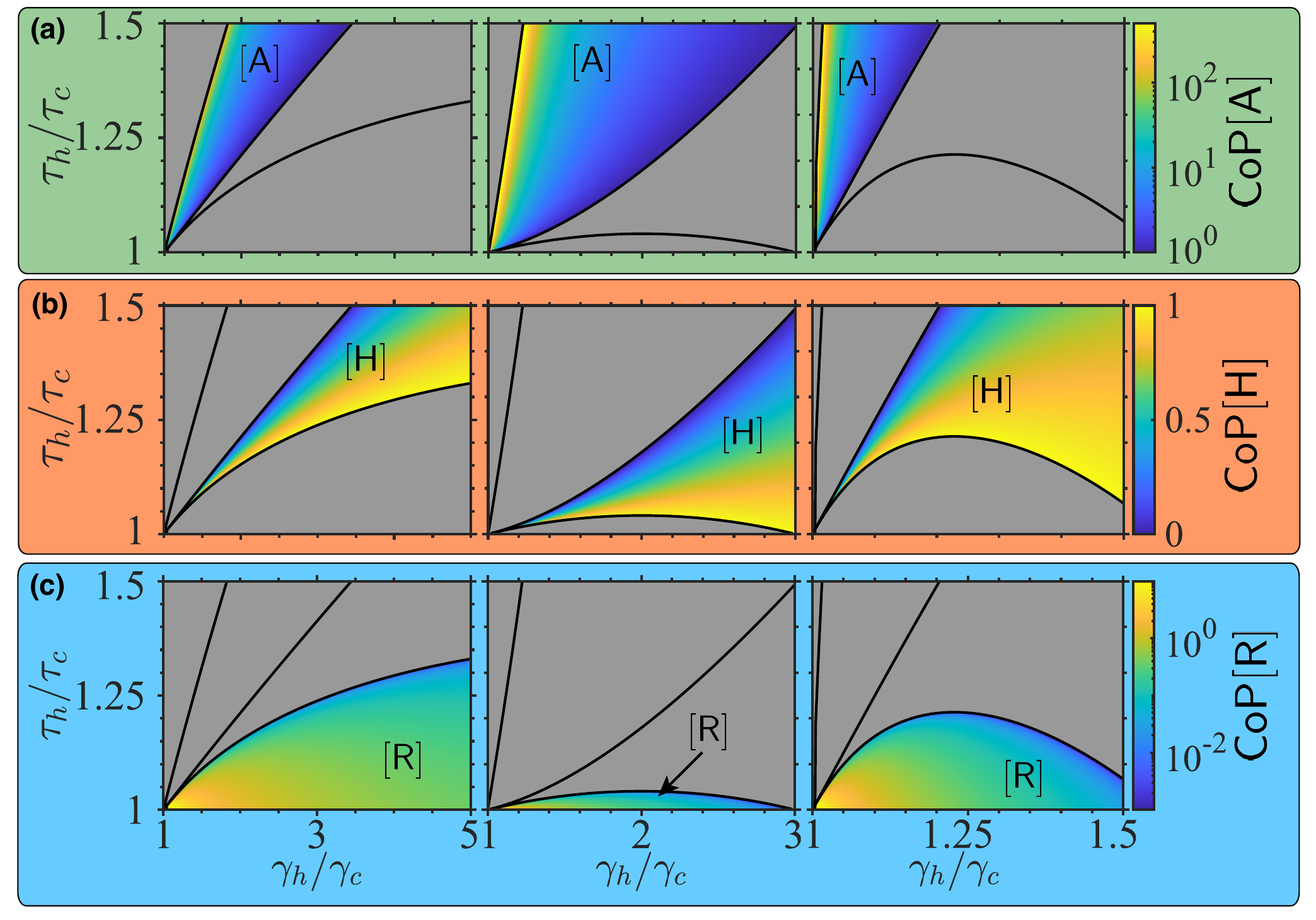}
    \caption{Coefficients of performance (CoP) of QTMs in the accelerator [A], heater [H], and refrigerator [R] regimes within the respective parameter ranges II, IV, and VI shown in Fig.~\ref{fig:operational_regimes}\,(e).}
\label{fig:CoP}
\end{figure}

The conditions of operating the Otto cycle as a thermal accelerator
are given by: $W\!>\!0$, $Q_1 \!>\! 0$, $Q_2\! >\! 0$, with $|Q_2|>|Q_1|$, where $W\!=\!0$ defines the border between the heat engine and the accelerator, whereas $Q_1\!=\!0$ defines the border between the accelerator and the next QTM, the heater; see Fig. \ref{fig:operational_regimes}, where we show the simplified schematics of all additional QTM's compared to the heat engine from Fig.~\ref{fig:Engine_Diagram}, which we repeat  here in panel (a) for comparison. The thermal accelerator \cite{buffoni2019quantum}, shown in panel (b) of Fig.  \ref{fig:operational_regimes}, enhances the natural flow of heat, taken into the working fluid from the hot reservoir, $Q_1$, and transferred to the cold reservoir, $Q_2$, which is aided by the working fluid through the investment of net mechanical work, $W>0$, in the process. According to Eq.~\eqref{eq:CoP}, its CoP is given by:
    \begin{equation}
        \mathrm{CoP[A]} = -\frac{Q_2}{W} = 1 + \frac{Q_1}{W} > 1,
    \end{equation}
where,
\begin{align}
    Q_2&= -\langle \hat{H} \rangle_h + \langle \hat{H} \rangle_c + \frac{N \hbar^2 \rho^2}{2m}(\gamma_h - \gamma_c) g^{(2)}_h(0),
\label{eq:Q2}\\
    Q_1 &= \langle \hat{H} \rangle_h - \langle \hat{H} \rangle_c - \frac{N \hbar^2 \rho^2}{2m}(\gamma_h - \gamma_c) g^{(2)}_c(0),
\label{eq:Q1}
\end{align}
whereas work $W$ is given by Eq.~\eqref{eq:Work_Uniform} as before. The magnitude of CoP[A] is shown in Fig.~\ref{fig:CoP}\,(a), where we note that, at the border between [E] and [A], the coefficient of performance diverges due to its inverse dependence on $W$.
Operation of this QTM, enabled through $g^{(2)}_h(0)\!<\!g^{(2)}_c(0)$, additionally requires that $|W_1|=\langle \hat{H} \rangle_{\mathbf{B}}\!-\!\langle \hat{H} \rangle_{\mathbf{C}}>0$ (where $W_1<0$) and $W_2=\langle \hat{H} \rangle_{\mathbf{A}}\!-\!\langle \hat{H} \rangle_{\mathbf{D}}>0$, which are proportional to $g^{(2)}_c(0)$ and $g^{(2)}_h(0)$, respectively, must individually remain smaller than the energy gap, $\Delta E \!=\!\langle \hat{H} \rangle_{\mathbf{B}}\!-\!\langle \hat{H} \rangle_{\mathbf{D}}$, between the hot and cold thermal states, as shown in the cycle diagram in Fig.~\ref{fig:operational_regimes}\,(b).
The conditions $|W_1|<\Delta E$ and $W_2<\Delta E$ may be expressed as 
\begin{align}
  N \frac{\hbar^2 \rho^2}{2m}(\gamma_h - \gamma_c) g^{(2)}_{h}(0) < \Delta E,\\
    N \frac{\hbar^2 \rho^2}{2m}(\gamma_h - \gamma_c) g^{(2)}_{c}(0) < \Delta E,
\end{align}
and are equivalent to $Q_1\!>\!0$ and $Q_2\!<\!0$, respectively.

\subsection{Heater [H]}

Operating the Otto cycle in the heater regime requires: $W > 0, Q_1 < 0, \, Q_2 < 0$, and is shown schematically in Fig.~\ref{fig:operational_regimes}\,(c). This QTM utilizes mechanical work to heat up both hot and cold thermal reservoirs. The border with the accelerator regime is defined by $Q_1\!=\!0$, and the lower border with the refrigerator QTM is defined by $Q_2\!=\!0$; see Fig.~\ref{fig:operational_regimes}\,(e). We note that the condition $Q_1\!<\!0$ corresponds to $W_2\!=\!\langle \hat{H} \rangle_{\mathbf{A}}\!-\!\langle \hat{H} \rangle_{\mathbf{D}}\!>\!\Delta E$, opposite to the condition for the accelerator regime [A].
Additionally, for a fixed temperature ratio, $\tau_h/\tau_c$, since an arbitrarily large interaction strength quench to $\gamma_h\gg 1$ incurs fermionization (i.e. $g^{(2)}_h(0)\!\to\!0$), operation as a heater is inevitable  as $\gamma_h/\gamma_c\to \infty$ at any fixed value of $\tau_h/\tau_c$ (see Appendix \ref{appendix:QTMs}).

If one considers the benefit of operation of the heater to be heating of \textit{both} reservoirs, then its CoP is trivially $\mathrm{CoP[H]}\!=\!-Q/W \!=\! 1$. Instead, we define the benefit of operation of this QTM to be the heating of the hot reservoir; thus 
 \begin{equation}\label{eq:cop_H}
         \mathrm{CoP[H]} = -\frac{Q_1}{W} = 1 - \frac{|Q_2|}{W} < 1,
    \end{equation}
which shown in Fig.~\ref{fig:CoP}\,(b). 

One may alternatively define the benefit of the heater as heating the cold reservoir, in which case the CoP would be given by:
\begin{equation}
         \mathrm{CoP[H]}' = -\frac{Q_2}{W} = 1 - \mathrm{CoP[H]} < 1.
    \end{equation}
Both definitions of $\mathrm{CoP[H]}$, however, are limited to be less than or equal to $1$ by energy conservation .

\subsection{Refrigerator [R]}

The conditions of operating the Otto cycle
as a refrigerator are: $W>0, \, Q_1 < 0, \, Q_2 > 0$, with $|Q_2|<|Q_1|$. The purpose of this thermal machine is to cool down the cold reservoir by extracting heat and dumping it into the hot reservoir, with the aid mechanical work done by the working fluid. The boundary between the refrigerator and the heater is defined by $Q_2\!=\!0$, see Fig.~\ref{fig:operational_regimes}\,(d).
The CoP for the refrigerator is given by \cite{SchroederD_ThermalPhysics}
    \begin{equation}
         \mathrm{CoP[R]} = \frac{Q_2}{W}=\frac{|Q_1|}{W}-1,
    \end{equation}
and is shown in Fig.~\ref{fig:CoP}\,(c); it diverges in the limit of infinitesimal quenches in both interaction strength and temperature, because the benefit of refrigeration, $Q_2$,  vanishes slower than the cost, $W$, in these limits (see Appendix \ref{appendix:QTMs}). Further, as noted in the section addressing the heater QTM, large interaction strength quenches inevitably incur operation as a heater for any fixed temperature ratio. This implies that refrigeration occurs only over a finite parameter region, most clearly visible in regimes IV and VI of Fig.~\ref{fig:operational_regimes}\,(e).

The role of the atom-atom local correlation function $g^{(2)}(0)$ in the thermal operation regimes and the respective boundaries of these additional QTM's is discussed further in Appendix \ref{appendix:QTMs}.

\section{Summary and outlook.}
\label{sec:summary}

We have proposed a sudden interaction-quench Otto cycle operating in a quantum many-body regime using a repulsive 1D Bose gas as a working fluid. Extracting net work from such a cycle in the heat engine regime was shown to be enabled by atom-atom correlations. Such correlations are characterized by Glauber's second-order correlation function, $g^{(2)}(0)$, which is a thermodynamic quantity that can be calculated from the exact TBA solution through application of the Hellmann-Feynman theorem to the Helmholtz free energy \cite{kheruntsyan2003pair}. Further, we have investigated other operational regimes of this Otto cycle, such as the thermal accelerator, heater, and refrigerator cycles, defining and examining their coefficients of performance.

Though our specific results for the net work and the efficiency were calculated for a uniform 1D Bose as an example, the broad conclusions arrived at here on the basis of Eqs.~\eqref{eq:Work} and \eqref{eq:Efficiency} are applicable to any other many-body system with short-range contact interactions and should aid the tests of quantum thermodynamic concepts and realization of novel QTMs in laboratory settings.

\section*{Acknowledgements}
 This work was supported through Australian Research Council Discovery Project Grant No. DP190101515.\\

~\\

\begin{appendix}

\noindent \textbf{\Large{Appendices}}\\

\noindent In these Appendices, we briefly review the Lieb-Liniger model of the one-dimensional (1D) Bose gas with contact interactions, and discuss how the interaction quench can be achieved via changes of the strength of the transverse confinement of the 1D Bose gas. We also discuss physical considerations for the sudden quench to be effectively instantaneous. We next present the relevant analytic results for the Glauber's second-order correlation $g^{(2)}(0)$ and the Hamiltonian energy $\langle \hat{H} \rangle$ in six different asymptotic regimes of the 1D Bose gas, used in the main text for evaluating the net work and efficiency of the Otto quantum heat engine. We also present equivalent numerical results obtained using the exact thermodynamic Bethe ansatz (TBA), and use these to explore a sudden quench Otto cycle operation  outside of the analytically tractable asymptotic regimes of the 1D Bose gas. This is followed by a discussion of the maximum efficiency at maximum work. Finally, we discuss the role of the pair correlation $g^{(2)}(0)$ in the operation of other quantum thermal machines (QTMs) presented in the main text, such as accelerator, heater, and refrigerator.

\section{The Lieb-Liniger model for the 1D Bose gas}
\label{appendix:LL}

The Lieb-Liniger model for the uniform 1D Bose gas with repulsive contact interactions  \cite{liebliniger} is described by the follwoing second-quantized Hamiltonian
\begin{equation}\label{eq:Hamiltonian}
\begin{split}
\begin{aligned}
    \hat{H} = \hat{H}^{kin} + \hat{H}^{int} 
    =- \frac{\hbar^2}{2m}\int dz \hat{\Psi}^\dagger \frac{\partial^2}{\partial z^2} \hat{\Psi}
    + \frac{\chi}{2} \int dz  \hat{\Psi}^\dagger \hat{\Psi}^\dagger \hat{\Psi} \hat{\Psi},
\end{aligned}
\end{split}
\end{equation}
where $m$ is the atomic mass, $\chi$ is the strength of the contact interactions (see main text), and $\hat{\Psi}^\dagger(z)$ and $\hat{\Psi}(z)$ are the bosonic field creation and annihilation operators, respectively. We additionally highlight the separation of the Lieb-Liniger Hamiltonian into its kinetic energy, $\hat{H}^{kin}$, and interaction energy, $\hat{H}^{int}$, components, to be referred to later.

Ground state solutions to this integrable model are dependent only on a single dimensionless interaction strength, $\gamma\!=\!m \chi/\hbar^2 \rho$, where $\rho\!=\!N/L$ is the linear density for $N$ particles in a system of size $L$. Finite temperature solutions, on the other hand, can be obtained using the Yang-Yang thermodynamic Bethe ansatz (TBA) \cite{yang1969thermodynamics}, and can be parameterized by an additional dimensional parameter, the dimensionless temperature $\tau \!=\! 2 m k_B T/\hbar^2 \rho^2$ \cite{kheruntsyan2003pair}.

\section{Transverse Otto cycle}
\label{appendix:Otto} 

For a magnetically trapped ultracold 1D Bose gas, the work done via transverse compression and expansion is ultimately magnetic: it is done by the magnetic field on the atomic dipole moments when $\omega_{\perp}$ is increased, or vice versa -- by the atomic dipole moments on the magnetic field when  $\omega_{\perp}$ is decreased. Alternatively, the change in the interaction strength $\chi$ is implemented through control over the $s$-wave scattering length $a_s$ via a magnetic Feshbach resonance \cite{chin2010feshbach}, in which case the nature of the work is still magnetic.

We use the term Otto cycle in the same sense as used to describe, e.g., a harmonic oscillator Otto engine \cite{Kosloff1984quantum,Kosloff2017quantum,li2018efficient,chen2019interaction,koch2022making,bouton2021quantum,zheng2014power,watanabe2017cycles,Abah_2017,Myers2022boosting}, wherein the harmonic oscillator frequency (rather than the volume of the system) is fixed as an external parameter during the thermalization strokes. In our case, it is the interaction strength that is fixed, which itself is proportional to the transverse harmonic confinement frequency of the 1D Bose gas.

\section{Instantaneity of the sudden quench}
\label{appendix:instantaneity} 

Realistically, a sudden quench of interaction strength from $\chi_{h(c)}$ to $\chi_{c(h)}$ would still occur over a finite duration $\Delta t$. The ``instantaneity'' of the quench utilized in the main text refers to the assumption that $\Delta t$ is much shorter than the characteristic time scale for longitudinal dynamics, i.e. that $\Delta t \!\ll\! m l_{\text{cor}}^2/\hbar$, where $l_{\text{cor}}$ is the characteristic short-range correlation length in the system, given, respectively, by: the healing length $l_h\!=\!\hbar/\sqrt{m \chi \rho}$ in regimes I and II;  thermal phase coherence length $l_\phi \!=\! \hbar^2 \rho / m k_B T$ in regime III; thermal de Broglie wavelength $\lambda_T \!=\! \sqrt{2 \pi \hbar^2 / m k_B T}$ in regime IV; absolute value of the 1D scattering length $|a_{1D}|=2\hbar^2/m\chi$ in the regime of high-temperature fermionization V; and the Fermi wavelength $\lambda_F \!=\! 2 / \rho$ in the Tonks-Girardeau regime of low-temperature fermionization VI. 
Thus, it is with respect to the \emph{longitudinal} dynamics that we refer to our quench as sudden. With respect to the \emph{transverse} dynamics, on the other hand, we are assuming that
$\Delta t$ is sufficiently long ($\Delta t\gg 2\pi/\omega_{\perp}$) compared to the characteristic transverse timescale, $2\pi/\omega_{\perp}$, governed by the transverse harmonic trap frequency $\omega_{\perp}$ \cite{Olshanii1:998}.
As a result, the quench would retain the system in the transverse ground state, and hence would not compromise the 1D character of the system. 
As such, the work done on (or by) the system during the unitary strokes can be regarded as transversely quasistatic.

\begin{figure*}[!tbp]
   \includegraphics[width=15cm]{ 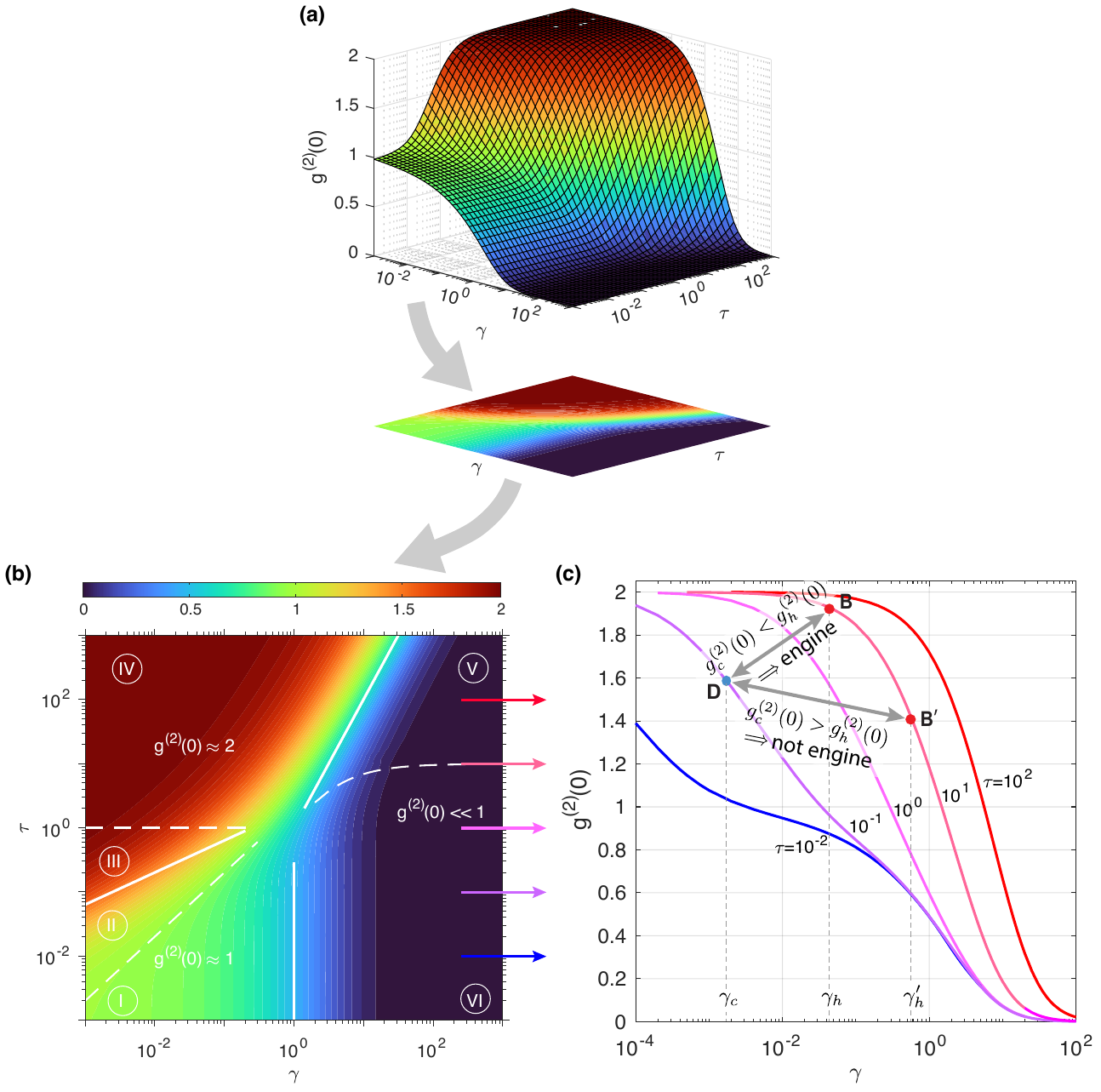}   
   \caption{Atom-atom correlations, described by Glauber's $g^{(2)}(0)$ correlation function, for the uniform 1D Bose gas evaluated using the exact Yang-Yang TBA \cite{kheruntsyan2003pair,yang1969thermodynamics}. Panel \textbf{a} shows $g^{(2)}(0)$ as a function of the dimensionless interaction strength, $\gamma$, and temperature, $\tau$. In panel \textbf{b} (same as Fig.~\ref{fig:regimes} of the main text), this is translated into a contour diagram, in which we also show the crossover boundaries (white solid and dashed lines) between the different asymptotic analytic regimes \cite{kheruntsyan2003pair}.
   Panel ($\textbf{c}$) shows line plots of $g^{(2)}(0)$ vs $\gamma$, at different fixed values of $\tau$, together with two possible choices, $\textbf{D-B}$ or $\textbf{D-B}'$, of the thermal equilibrium operating points of the Otto cycle from Fig.~\ref{fig:Engine_Diagram}; 
   as we see, according to Eq.~\eqref{eq:Work_Uniform}, 
   operating the Otto cycle as an engine (with $W<0$) can be achieved between the points $\textbf{D-B}$ ($\gamma_c\longleftrightarrow \gamma_h$), where $g^{(2)}_c(0)\!<\!g^{(2)}_h(0)$, but not between $\textbf{D-B}'$ ($\gamma_c\longleftrightarrow \gamma'_h$), where $g^{(2)}_c(0)\!>\!g^{(2)}_h(0)$ due to the stronger interaction quench, even though the temperature at $\textbf{D}$ is still lower than at $\textbf{B}'$.
   }
  \label{fig:g2_parameter_space}
\end{figure*}

\section{Glauber's second order correlation function}
\label{appendix:Glauber} 

The two-point correlation function may be generally defined through
\begin{equation}\label{eq:g2}
    g^{(2)}(z,z') = \frac{\langle \hat{\Psi}^\dagger(z)\hat{\Psi}^\dagger(z')\hat{\Psi}(z')\hat{\Psi}(z) \rangle}{\rho(z)\rho(z')}.
\end{equation}
For a uniform (translationally invariant) system with $\rho(z')\!=\!\rho(z)\!=\!\rho$, this $g^{(2)}(z,z')$ depends only on the relative distance $|z-z'|$, i.e. $g^{(2)}(z,z')\!=\!g^{(2)}(|z-z'|)$. If one is interested in the same point ($z\!=\!z'$) correlation function, as utilized in the main text for calculation of the net work and efficiency of the quantum Otto cycle, this in turn becomes $g^{(2)}(0)$.

The 1D Bose gas can be characterized by six distinct asymptotic regimes defined through the same-point correlation function \cite{kheruntsyan2003pair}, as shown in Fig.~\ref{fig:g2_parameter_space}. 
The weakly interacting ($\gamma\ll 1$), low temperature ($\tau \ll 2\sqrt{\gamma}$) quasicondensate regime can be treated using the Bogoliubov theory for quasicondensates  \cite{Mora-Castin-2003}, and is characterised by suppressed density fluctuations, but fluctuating phase. This may be subdivided into regions dominated by quantum (I) and thermal (II) fluctuations \cite{kheruntsyan2003pair}. At higher temperatures, the gas becomes nearly ideal, and can be treated using perturbation theory with respect to $\gamma$ \cite{kheruntsyan2003pair,Sykes2008}. This asymptotic region may in turn be subdivided into quantum degenerate (III) and non-degenerate (IV) regimes. Finally, in the strongly interacting regime ($\gamma \gg 1$), where the Fermi-Bose gas mapping applies, the 1D Bose gas can be well approximated by a nearly ideal Fermi gas, and can be treated using perturbation theory with respect to $1/\gamma$ \cite{kheruntsyan2003pair,Sykes2008}.
This regime can be further subdivided into two regions corresponding to high-temperature (V) and low-temperature (VI) fermionization.

In each of these asymptotic regimes, the pair correlation function $g^{(2)}(0)$ can be derived in closed approximate analytic form, where we additionally define the boundary of these regimes in terms of $\gamma$ and $\tau$:
\begin{align}
    \label{eq:g2_I}
    \text{I}\!:\, &g^{(2)}(0) \!\simeq\! 1 - \frac2{\pi} \gamma^{1/2} +  \frac{\pi\, \tau^2}{24\, \gamma^{3/2}},   
    \,\left[\frac{\tau}{2}\!\ll\!\gamma\ll 1\right],\\
    \label{eq:g2_II}
    \text{II}\!:\,\,  &g^{(2)}(0) \simeq 1 + \frac{\tau}{2\sqrt{\gamma}}\,\,,
    \quad \left[2\gamma\ll \tau\ll 2\sqrt{\gamma}\right],\\
    \label{eq:g2_III}
    \text{III}\!:\,\, &g^{(2)}(0) \simeq 2 - \frac{4\gamma}{\tau^2},
     \quad \left[2\sqrt{\gamma} \ll \tau \ll 1\right], \\
    \label{eq:g2_IV}
    \text{IV}\!:\,\, &g^{(2)}(0) \simeq 2 - \gamma\sqrt{\frac{2\pi}{\tau}},
    \,\,\,\left[\tau \gg \text{max}\{1,\gamma^2\}\right],\\
    \label{eq:g2_V}
     \text{V}\!:\,\, &g^{(2)}(0) \simeq \frac{2\tau}{\gamma^2},
     \,\,\left[\frac{\pi^2}{(1+2/\gamma)^2} \ll \tau \ll \gamma^2\right],\\
\label{eq:g2_VI}    \text{VI}\!:\, &g^{(2)}(0) \simeq \frac{4\pi^2}{3\gamma^2}\left(1\! +
\!\frac{\tau^2}{4\pi^2}\right)\!,\left[\tau\!\ll\! \frac{\pi^2}{(1+2/\gamma)^2},\, \gamma \!\gg\! 1\right]\!. 
\end{align}

Further, we may express the total energy of system in each asymptotic regime as \cite{DeRosi2022anomaly},
\begin{align}
    \label{eq:H_I}
    \text{I}:\,\,\, & \langle \hat{H} \rangle \!\simeq\! N\frac{\hbar^2 \rho^2}{2m}\left( \gamma - \frac{4}{3\pi}\gamma^{3/2} + \frac{\pi}{12}\frac{\tau^2}{\gamma^{1/2}} \right),\\
    \label{eq:H_II}
     \text{II}:\,\,\, & \langle \hat{H} \rangle \!\simeq\! N\frac{\hbar^2 \rho^2}{2m}\left(\gamma + \frac{\zeta(3/2)}{4\sqrt{\pi}}\tau^{3/2} + \frac{\zeta(1/2)}{2\sqrt{\pi}}\tau^{1/2}\gamma \right),\\
    \label{eq:H_III}
     \text{III}:\,\,\, & \langle \hat{H} \rangle \!\simeq\! N\frac{\hbar^2 \rho^2}{2m}\left( \frac{1}{2}\zeta(3/2) + 2 \gamma -  \frac{6\gamma^2}{ \tau^2} \right),\\
    \label{eq:H_IV}
     \text{IV}:\,\,\, & \langle \hat{H} \rangle \!\simeq\! N\frac{\hbar^2 \rho^2}{2m}\left( \frac{\tau}{2}  + 2 \gamma - \frac{3}{2} \sqrt{\frac{\pi}{2}}\frac{\gamma^2}{\tau^{1/2}}\right),
\end{align}
\begin{align}
    \label{eq:H_V}
      \text{V}:\,\,\, & \langle \hat{H} \rangle \!\simeq\! N\frac{\hbar^2 \rho^2}{2m}\left( \frac{\tau}{2} + \frac{1}{2} \sqrt{\frac{\pi}{2}}\tau^{1/2} - \sqrt{\frac{\pi}{2}} \frac{\tau^{1/2}}{ \gamma} \right),\\
    \label{eq:H_VI}
     \text{VI}:\,\,\, & \langle \hat{H} \rangle \!\simeq\! N\frac{\hbar^2 \rho^2}{2m}\left( \frac{\pi^2}{3} - \frac{4 \pi^2}{3 \gamma} +\frac{\tau^2}{3\gamma} \right),
\end{align}
where $\zeta(s)$ is the Riemann zeta function of $s \!\in\! \mathbb{R}$.

As described in the main text, the normalized local second-order correlation function may be rearranged and integrated for the total (integrated) correlation function,
\begin{equation}\label{eq:G2}
\begin{split}
\begin{aligned}\!
    \overline{G^{(2)}} \equiv \! \int dz \langle \hat{\Psi}^\dagger(z)\hat{\Psi}^\dagger(z)\hat{\Psi}(z)\hat{\Psi}(z) \rangle \!=\! \int_0^L \!dz g^{(2)}(0) \rho^2.
\end{aligned}
\end{split}
\end{equation}
Utilizing the linear density, $\rho \!=\!N/L$, this may be expressed as $\overline{G^{(2)}}\!=\! N \rho g^{(2)}(0)$, which is used in Eq.~\eqref{eq:Work_Uniform} 
of the main text for expressing the exact net work of the uniform 1D Bose gas in terms of $g^{(2)}(0)$.

\begin{figure}[t!]
      \centering
    \includegraphics[width=12cm]{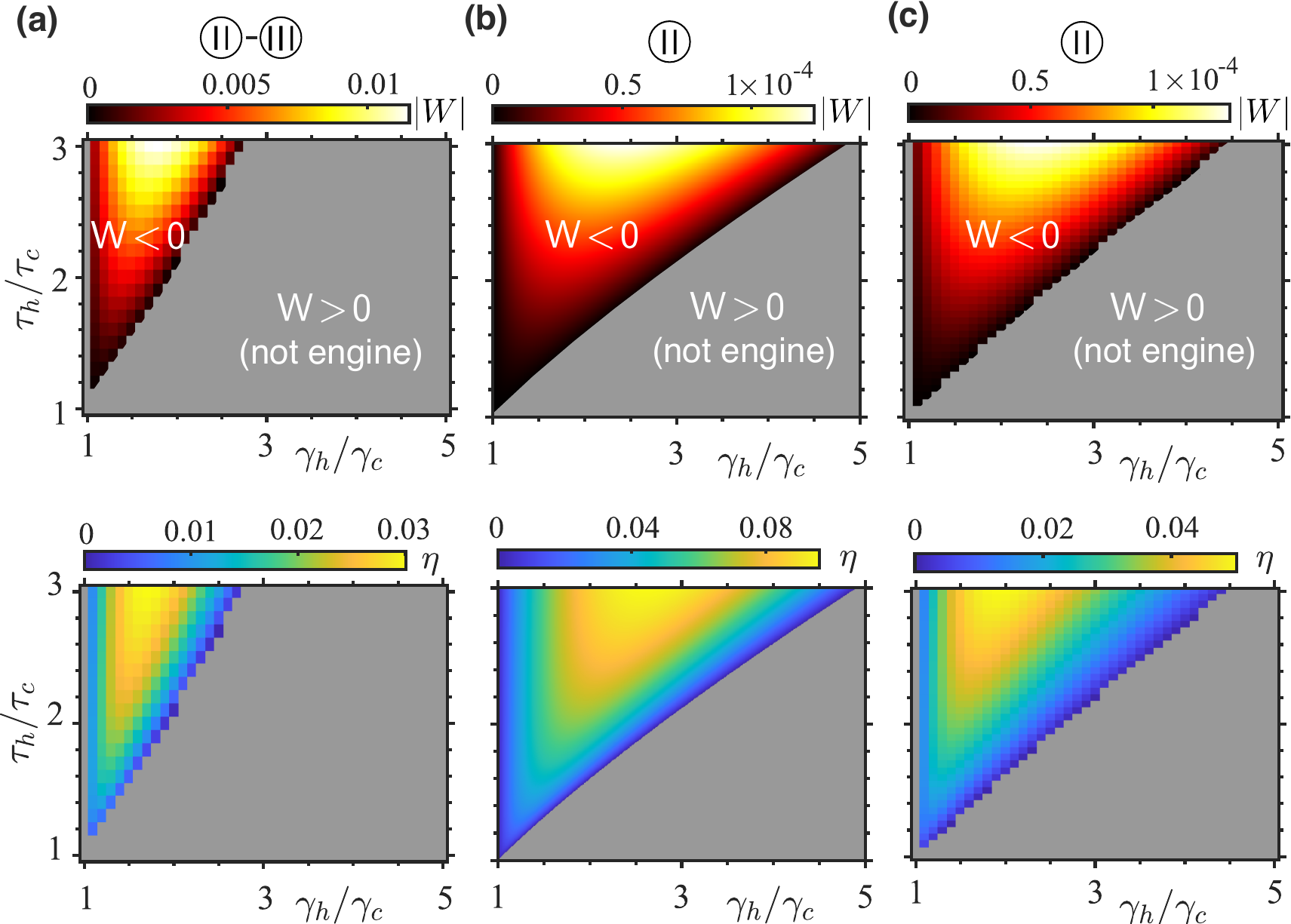}
    \caption{Performance of the sudden interaction quench quantum Otto cycle, numerically evaluated via the thermodynamic Bethe ansatz. Panel \textbf{a} demonstrates numerically evaluated net work and efficiency for a system with a cold thermal state defined by $\gamma_c \!=\! 0.1$, $\tau_c \!=\! 0.5$, lying on the border of regimes II and III (see Fig.~\ref{fig:g2_parameter_space}), and thus lying outside the range of the analytic approximations utilized in the main text. Panel (b) is a copy of Fig.~\ref{fig:engine_performance}\,(a) of the main text 
    shown here for comparison with the results of numerical TBA evaluation of the same cycle shown in panel (c). Here, there is excellent agreement in the net work between panels (b) and (c), with small disagreement under large interaction strength and temperature ratios, as the hot thermal state is approaching the edge of the asymptotic regime where the analytic approximations become less applicable.}
\label{fig:Numeric_Comparison}
\end{figure}

\section{Exact thermodynamic Bethe ansatz results}
\label{appendix:TBA} 

Experimental realization of a 1D Bose gas often falls outside the asymptotic regimes where analytic approximations are applicable. In such situations, we may utilize the exact Yang-Yang thermodynamic Bethe ansatz \cite{yang1969thermodynamics,korepin1993} to evaluate the equilibrium properties of the gas required for calculating net work and efficiency via Eqs.~\eqref{eq:Work} and \eqref{eq:Efficiency} 
of the main text, respectively. This is presented in Fig.~\ref{fig:Numeric_Comparison}\,(a) for experimentally realistic set of system parameters that inhabit the boundary between asymptotic parameter regimes II and III (see Fig.~\ref{fig:g2_parameter_space})\,(b). 
Further, one may utilize the exact TBA to confirm the results derived via approximate analytics in the main text. This is illustrated in  Figs.~\ref{fig:Numeric_Comparison}\,(b) and (c), where we see excellent agreement between these results when the parameters $\gamma$ and $\tau$ are sufficiently deep into the analytic asymptotic regimes.

\section{Maximum efficiency and maximum work}
\label{appendix:max-efficiency} 

For a fixed ratio of temperatures, it was noted in the main text that the interaction strength ratio corresponding to maximum work approximately coincides with that for maximum efficiency, which is uncommon for highly nonequilibrium engine cycles \cite{jaramillo2016quantum,Kosloff2017quantum}. 
In the sudden interaction-quench Otto cycle, such coincidence occurs due to the dependence of the total energy of the hot and cold thermal equilibrium states on the interaction strength.

As shown in Eq.~\eqref{eq:Hamiltonian}, the total energy may be separated into its kinetic energy, which scales predominately with temperature, and interaction energy, which scales predominately with interaction strength.
Thus, for a fixed ratio of temperatures, $\tau_h/\tau_c$, the difference between the total energies of the hot and cold thermal state may be given as a sum of two terms: the first is the kinetic energy difference, determined by the temperature ratio and therefore approximately constant, the second given by the interaction energy difference, which scales with the interaction strengths, $\gamma_h$ and $\gamma_c$, of the hot and cold thermal states as
\begin{equation}\label{eq:interaction_energy_diff}
    \langle \hat{H}^{int} \rangle_h \!-\! \langle \hat{H}^{int} \rangle_c \!=\! N \frac{\hbar^2 \rho^2}{2m}\left(\gamma_h g^{(2)}_h(0) \!-\!\gamma_c g^{(2)}_c(0)\right).
\end{equation}

However, when operating within a single asymptotic regime under a moderate quench of interaction strength, the $g^{(2)}(0)$ correlation function is only slowly varying with $\gamma$. This means, to first approximation, $g^{(2)}_h(0) \!\simeq\!g^{(2)}_c(0)$, which in turn transforms the interaction energy difference to
\begin{equation}
    \langle \hat{H}^{int} \rangle_h \!-\! \langle \hat{H}^{int} \rangle_c \!\simeq\! N \frac{\hbar^2 \rho^2}{2m}\left(\gamma_h  \!-\!\gamma_c \right) g^{(2)}_c(0).
\end{equation}
The heat intake, which is given by Eq.~\eqref{eq:Q1} of the main text, 
is therefore well approximated by
\begin{align}
    Q_1 = \langle \hat{H} \rangle_h - \langle \hat{H} \rangle_c - \frac{N \hbar^2 \rho^2}{2m}(\gamma_h - \gamma_c) g^{(2)}_c(0) 
    \simeq \langle\hat{H}^{kin} \rangle_h - \langle \hat{H}^{kin} \rangle_c,
\end{align}
which is approximately constant for a fixed temperature ratio, as detailed above. Therefore, the efficiency, which is given by $\eta \!=\!W/Q_1$, scales predominately with $W$, hence $\eta\!\propto\!W$.

\begin{figure*}[!tbp]
    \centering
    \includegraphics[width=12cm]{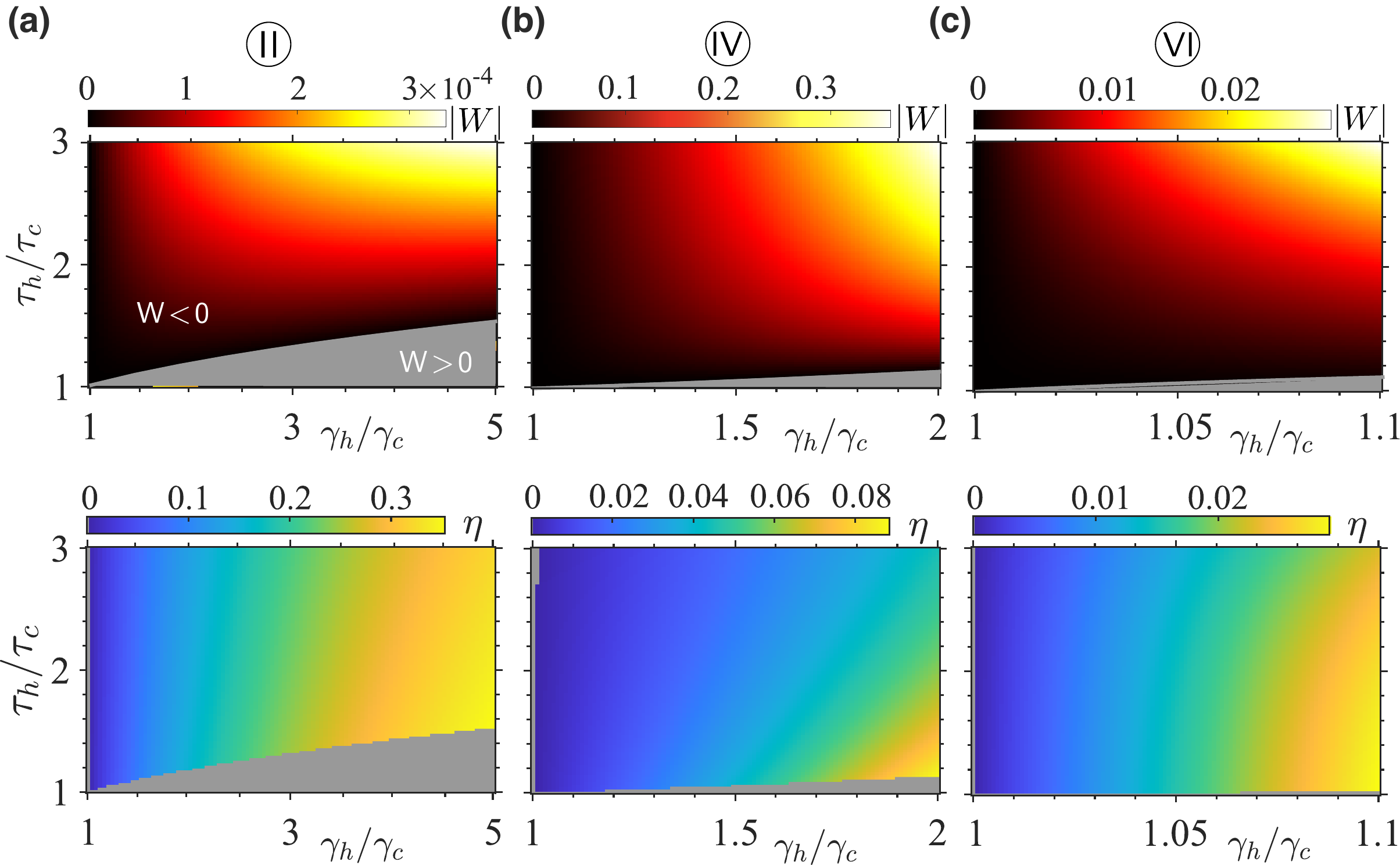}
    \caption{Performance of the isentropic interaction-driven quantum Otto cycle, for the same parameters as in the sudden quench scenarios of Figs.~\ref{fig:engine_performance}\,\textbf{a}--\textbf{c}. As one approaches the boundary of the engine regime, defined by $W\!=\!0$, the efficiency, which is given by the ratio of the net work, $W$, to the heat intake, $Q_1$, remains approximately unchanged, unlike the $W$ itself which approaches 0 near the boundary. However, as the efficiency is only a valid metric in the engine regime, it is set to gray scale outside the engine regime.
 }
\label{fig:SQ_AD}
\end{figure*}

\section{Isentropic quench}
\label{appendix:isentropic-quench}

The conventional isentropic Otto engine cycle is, by definition, fully reversible, and as such obtains the maximum possible values for both net work and efficiency \cite{SchroederD_ThermalPhysics}. The isentropic interaction-driven Otto cycle for the uniform 1D Bose gas was introduced in Ref.~\cite{chen2019interaction}, where they employed the Tomonaga-Luttinger liquid (TLL) theory to derive approximate analytic results for the net work and efficiency. The TLL theory captures the low energy regimes of the uniform 1D Bose gas \cite{1D_entropy_2017}, corresponding to regions I and VI of Fig.~\ref{fig:g2_parameter_space}.

Here, we analyze the performance of the isentropic Otto engine cycle under higher temperatures, i.e., outside of the regime of applicability of the TLL theory. This can be done by 
utilizing the conservation of entropy condition for the work strokes $W_1$ and $W_2$ of Fig.~\ref{fig:Engine_Diagram} \cite{chen2019interaction}, along with the TBA or approximate analytic results for entropy available in the uniform 1D Bose gas \cite{DeRosi2022anomaly},
\begin{align}
    \label{eq:S_I}
    \text{I}:\,\,\, &  S  \!\simeq\! N k_B\left( \frac{\pi}{6}\frac{\tau}{\sqrt{\gamma}}  - \frac{\pi^3}{240}\frac{\tau^3}{\gamma^{5/2}} \right),\\
    \label{eq:S_II}
     \text{II}:\,\,\, &  S  \!\simeq\!  N k_B\left( \frac{3 \zeta(3/2)}{4 \sqrt{\pi}} \sqrt{\tau} - \sqrt{\gamma} - \frac{ \zeta(1/2)}{2 \sqrt{\pi}} \frac{\gamma}{\sqrt{\tau}} \right),\\
    \label{eq:S_III}
     \text{III}:\,\,\, &  S  \!\simeq\!  N k_B\left( \frac{3 \zeta(3/2)}{ 4 \sqrt{\pi}} \sqrt{\tau} - \frac{4 \gamma^2}{
    \tau^3} \right),\\
      \label{eq:S_IV}
     \text{IV}:\,\,\, &  S  \!\simeq\!  N k_B\left( \ln \left( \frac{\sqrt{\tau}}{2 \sqrt{\pi}}\right) + \frac{3}{2} + \frac{\sqrt{\pi}}{\sqrt{2 \tau}} - \frac{1}{2} \sqrt{\frac{\pi}{2 \tau^3}} \gamma^2 \right),\\
    \label{eq:S_V}
      \text{V}:\,\,\, &  S  \!\simeq\!  N k_B\left(\ln \left( \frac{\sqrt{\tau}}{2 \sqrt{\pi}}\right) + \frac{3}{2} + \frac{\sqrt{\pi}}{\sqrt{2 \tau}} + \frac{2}{\gamma} + \frac{\sqrt{\pi}}{\gamma \sqrt{2\tau}} \right),\\
    \label{eq:S_VI}
    \text{VI}:\,\,\, &  S  \!\simeq\!  N k_B\left( \frac{\tau}{6} + \frac{\tau^3}{45 \pi^2} + \frac{2 \tau}{3 \gamma } + \frac{2 \tau}{3 \gamma^2 }\right).
\end{align}

In Fig.~\ref{fig:SQ_AD} we demonstrate performance of the isentropic engine cycle (using the analytic results), which may be directly compared against that of the sudden quench engine cycles explored in Fig.~\ref{fig:engine_performance}. 
The difference in the parameter range over which this cycle operates as an engine, when compared with the sudden quench engine cycle, may be attributed to the fact that this is a fundamentally different protocol, and thus is not expected to operate within the same bounds as its sudden quench equivalent.
Notably, over the range of parameters that the sudden quench Otto cycle operates as an engine, its performance -- while being inferior to that of the ideal, isentropic Otto cycle as expected -- remains on the same order of magnitude in both net work and efficiency, despite being highly nonequilibrium and thus irreversible. 

~

\section{Thermal operation regimes of other QTM's}
\label{appendix:QTMs} 

Under large interaction strength quenches, for fixed temperatures $\tau_c$ and $\tau_h$, it was noted in the main text that the heater is the inevitable mode of operation. This scenario requires the fulfilment of two conditions: first, we require $Q_1\!<\!0$, meaning the magnitude of the work out, $|W_1|\!=\!N\frac{\hbar^2 \rho^2}{2m}(\gamma_h\!-\!\gamma_c)g^{(2)}_c(0)$ (where $W_1<0$) exceeds the energy gap between the hot and cold thermal states, given by $\langle \hat{H} \rangle_h \!-\!\langle \hat{H} \rangle_c$. We further require $Q_2\!>\!0$, meaning that the work in, $W_2\!=\!N\frac{\hbar^2 \rho^2}{2m}(\gamma_h\!-\!\gamma_c)g^{(2)}_h(0)$ (where $W_2>0$), must remain less than this same gap (see Fig.~\ref{fig:operational_regimes}\,(c) of the main text).

As detailed above, in Appendix \ref{appendix:max-efficiency} on maximum efficiency at maximum work, the total energy difference between the hot and cold thermal states for fixed temperature ratios, $\tau_h/\tau_c$, is given by a sum of the kinetic energy difference, which is approximately constant, and the interaction energy difference, where the correlation function is approximately constant in a single asymptotic regime, and hence given by Eq.~\eqref{eq:interaction_energy_diff}.
In contrast, for a \textit{large} quench in interaction strength, the correlation function is no longer approximately constant, and $g^{(2)}_h(0)$ is strongly monotonically decreasing as a function of $\gamma_h$, i.e. $g^{(2)}_h(0) \!<\!g^{(2)}_c(0)$. We therefore find that the work input, $W_2$, exceeds the interaction energy difference given in Eq.~\eqref{eq:interaction_energy_diff},
\begin{equation}
    W_2 \propto (\gamma_h - \gamma_c) g^{(2)}_c(0) > \gamma_h g^{(2)}_h(0) \!-\!\gamma_c g^{(2)}_c(0).
\end{equation}
Further, as the kinetic energy term is approximately constant, $W_2$ inevitably exceeds the difference in total energy between the hot and cold thermal states due to its linear dependence on $\gamma_h$.

Similarly, since $g^{(2)}_h(0)$ monotonically decreases with $\gamma_h$ for large quenches of interaction strength, the magnitude of the work output, $|W_1| \propto (\gamma_h - \gamma_c) g^{(2)}_h(0)$, remains less than the energy gap between the hot and cold thermal states:
\begin{equation}
    |W_1| \propto (\gamma_h - \gamma_c) g^{(2)}_h(0) < \gamma_h g^{(2)}_h(0) \!-\!\gamma_c g^{(2)}_c(0).
\end{equation}
These two conditions, taken together, imply operation as a heater under large interaction quenches.

In contrast, for any fixed value of $\gamma_c$ and $\gamma_h$, an increasingly higher temperature of the hot thermal state, $\tau_h$, means that the corresponding correlation function, $g^{(2)}_h(0)$, monotonically increases towards its maximum value of $g^{(2)}_h(0) \!\simeq\!2$, which is achieved in regime IV, defined in Eq.~\eqref{eq:g2_IV}. Thus, there is always a value of $\tau_h$ such that $g^{(2)}_h(0) \!>\!g^{(2)}_c(0)$, turning the Otto cycle into the engine operation regime.

Finally, in the refrigerator thermal operation regime, for $\tau_h=\tau_c$ and an infinitesimal quench of interaction strength, $\gamma_h-\gamma_c=\delta \gamma$, the net work vanishes as $W\!\propto  (\gamma_h -\gamma_c)$ $\times(g^{(2)}_h(0)-g^{(2)}_c(0))\propto\delta \gamma^2$. This occurs as the zeroth order terms in the correlation function cancel when taking their difference in a single asymptotic regime. In contrast, the heat intake, $Q_1$, depends on the difference in total energy, which to first order vanishes as $Q_1\!\propto\!\delta \gamma$. This results in $\mathrm{CoP[R]}\!=\!|Q_1|/W\!-\!1\!\propto\delta\gamma^{-1}$, which diverges as $\delta \gamma \!\to\! 0$, as noted in the main text.

\end{appendix}


\nolinenumbers

\end{document}